\documentstyle[sprocl,12pt,psfig]{article} 
 
\catcode`\@=11 
 
\def\mycite{\@ifnextchar [{\@tempswatrue\@mycitex}{\@tempswafalse\@mycitex[]}} 
\def\mcite{\@ifnextchar [{\@tempswatrue\@mycitex}{\@tempswafalse\@mycitex[]}} 
 
\def\@mycitex[#1]#2{\if@filesw\immediate\write\@auxout{\string\citation{#2}}\fi 
 \def\@citea{}\@mycite{\@for\@citeb:=#2\do 
    {\@citea\def\@citea{,\penalty\@m\ }\@ifundefined 
       {b@\@citeb}{{\bf ?}\@warning 
       {Citation `\@citeb' on page \thepage \space undefined}}%
\hbox{\csname b@\@citeb\endcsname}}}{#1}}

\def\@mycite#1{[{#1}]} 
 
\catcode`\@=12 
\def\be{\begin{equation}}
\def\ee{\end{equation}}
\def\bea{\begin{eqnarray}}
\def\eea{\end{eqnarray}}

\begin{document}
\begin{flushright} 
NBI--HE--98--27\\ 
MPS--RR--1998-25 \\
ITEP--TH--50/98
\end{flushright} 
\vspace{0.4in}

\title{Thermodynamics of D0-branes in matrix theory~\footnote{
This work is supported in part by NATO Grant CRG 970561, by NSERC and 
by MaPhySto -- Centre for Mathematical Physics and Stochastics.}} 
\author{J. Ambj{\o}rn} 
\address{Niels Bohr Institute\\Blegdamsvej 17\\Copenhagen, 2100 Denmark\\
\tt ambjorn@nbi.dk}
\author{Y. M. Makeenko}
\address{Institute of Theoretical and Experimental 
Physics\\ B. Cheremushkinskaya 25 \\
Moscow, 117218 Russian Federation\\ \tt makeenko@itep.ru }
\author{G. W. Semenoff} 
\address{Department of Physics and Astronomy
\\University of British Columbia \\6224 
Agricultural Road\\Vancouver, British Columbia V6T 1Z1, Canada\\
\tt semenoff@physics.ubc.ca}

\maketitle

\vspace{0.2in}

\abstracts{
We examine the matrix theory representation of D0-brane dynamics at finite
temperature.  In
this case, violation of supersymmetry by temperature leads to a
non-trivial static potential between D0-branes at any finite
temperature. We compute the static potential in the 1-loop
approximation and show that it is short-ranged and attractive.  
We compare the result with the computations in superstring theory.
We show that thermal states of D0-branes can be reproduced by matrix
theory only when certain care is taken in integration over the moduli
space of classical solutions in compactified time.} 
\newpage\setcounter{page}{1}

\setcounter{equation}{0} 

\setcounter{page}{2}

\section{Introduction}

\begin{figure}
\vspace{-0.3in} 
\hspace{1.5in}
\psfig{figure=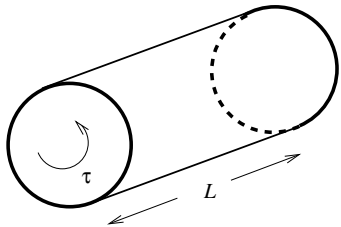,height=1.5in} 
\vspace{-0.3in} 
\caption[x]{\footnotesize Annulus diagram for D0-brane interactions.
The bold lines represent world-lines of the D0-branes, separated
by the distance $L$, which go along the periodic temporal direction. 
They bound the string world-sheet.} 
\label{f:annulus} 
\end{figure}

Dirichlet p-branes are p+1-dimensional hypersurfaces on which
superstrings can begin and end (see~\cite{Pol96,wati} for a review).
The low energy dynamics of an ensemble of N parallel Dp-branes can be
described by the U(N) supersymmetric gauge theory obtained by
dimensional reduction of ten dimensional supersymmetric Yang-Mills
theory to the p+1-dimensional world-volume of the brane.\cite{witt2}
The Yang-Mills theory gives an accurate perturbative representation of
the Dp-brane dynamics when the separations between the branes is
large.\cite{gab,kp,dkps} It represents a truncation of the full string
spectrum to the lowest energy modes.  The full string theoretical
interactions between a pair of Dp-branes is computed by considering
the annulus diagram shown in fig.~\ref{f:annulus}.  The short distance
asymptotics of this diagram are dominated by the open string sector
whose lowest modes are the fields of ten dimensional supersymmetric
Yang-Mills theory.  On the other hand, long distance asymptotics are
most conveniently described by the dual description of this diagram as
a closed string exchange and the relevant field theoretical modes are
those of ten dimensional supergravity.  That these are also
represented by the dimensionally reduced super Yang-Mills theory is a
result of supersymmetry and the fact that, for fixed Dp-brane
positions, the ground state is a BPS state.  At zero temperature,
because of supersymmetry, the interaction potential between a pair of
static D0-branes vanishes independently of their separation.  Their
effective action has been computed in an expansion in their
velocities, divided by powers of the separation and is known to
be~\cite{b,dkps}
\begin{equation}
S_{\rm eff}(T=0)= \int dt \left( 
\frac{1}{2g_s\sqrt{\alpha'}} \sum_{\alpha=1}^2 ( {\dot {\vec q}}\,{}^\alpha 
) ^2 - 
\frac{15}{16}\left( \alpha'\right)^3
\frac{\vert {\dot{\vec q}}\,{}^1-{\dot{\vec q}}\,{}^2
\vert^4}{\vert {\vec q}\,{}^1-{\vec q}\,{}^2 \vert^7}+\ldots \right) 
\label{T=0}
\end{equation}
This result agrees with the effective potential for the interaction 
of D0-branes in ten dimensional supergravity. Note that, for weak string 
coupling, the D0-brane is very heavy.

In this paper, we shall consider the description of D0-brane
interactions in type IIA superstring theory using matrices.  Even at
very low temperatures, non-BPS states are important to the leading
temperature dependence.
We perform 1-loop computation of the effective interaction between
static D0-branes in the matrix theory at finite temperature and
compare with the known superstring computations.  We show that the
results of the two computations are similar in the low temperature
limit but an extra integration over 
 the temporal component of the gauge
field, is present in the matrix
theory. At finite temperature, because the Euclidean 
time is compact, the  temporal gauge field  can not be removed  
by a gauge transformation.  {\it This integration
is needed in order to describe correctly thermodynamics of $D0$-branes
both in the matrix and superstring theories}.

The paper is organized as follows.  In section~2 we discuss the
formulation of the matrix theory at finite temperature. In section~3
we perform one loop computation of the effective interaction between
static D0-branes at finite temperature and show that it is 
attractive, and short-ranged. In section~4 we compare this result with
the superstring computations and discuss the conditions under which
the two computations agree. Section~5 is devoted to the discussion of
our results and, in particular, the origin of the divergence of the
classical thermal partition function of D0-branes which is cured by
quantum statistics.

\section{Matrix theory at finite temperature}

We shall consider the matrix theory description~\cite{bfss} of the effective
dynamics of D0-branes in a type-IIA superstring theory which is
derived by the reduction of ten dimensional supersymmertic 
Yang-Mills theory which has the action
\begin{equation}
S_{\rm YM}[A,\theta]=\frac{1}{g_{YM}^2}\int d\tau {\rm TR}\left(
\frac{1}{4}F_{\mu\nu}^2+\frac{i}{2}\theta\gamma_\mu D_\mu\theta\right)
\end{equation}
to zero spatial dimension: $A_\mu=A_\mu(\tau)$, $\theta=\theta(\tau)$.

The thermal partition function of this theory is given by
\begin{equation}
Z_{\rm YM}=\int[dA(\tau)][d\theta(\tau)]\exp\left( -S_{\rm YM}[A,\theta]\right)
\label{thermal}
\end{equation}
where $S_{\rm YM}$ is the Euclidean action and the time coordinate is
periodic. The bosonic and fermionic coordinates have periodic and
anti-periodic boundary conditions,
\begin{eqnarray}
A_\mu(\tau+\beta)= A_\mu(\tau),   \\
\theta(\tau+\beta)=-\theta(\tau),\\
\beta=1/k_BT ,
\label{b.c}
\end{eqnarray}
where $T$ is the temperature and $k_B$ is Boltzmann's constant.  Gauge
fixing will be necessary and will involve introducing ghost fields
which will have periodic boundary conditions.

The representation (\ref{thermal}) of the  thermal partition 
function can be derived in the standard way starting from the known
Hamiltonian of the matrix theory~\cite{bfss} and representing
the  thermal partition function 
\begin{equation}
Z_{\rm YM}= {\rm tr}\, e^{-\beta H}  
\end{equation}
via the path integral. The trace is calculated here over all states
 obeying Gauss's law which is taken care by the integration
over $A_0$ in (\ref{thermal}). This representation of the  matrix theory
at finite temperature have been already discussed \cite{OZ98,MOP98,Sat98},
but the temperature induced interaction between $D0$-branes described
below was never identified. 

In matrix theory, the diagonal components of the gauge fields,
$\vec{a}^\alpha\equiv \vec{A}^{\alpha\alpha}$, are interpreted as the
position coordinates of the $\alpha$-th D0-brane and they should be
treated as collective variables. Static configurations play a special
role since they satisfy classical equations of motion 
with the periodic boundary conditions and dominate
the path integral as $g^2_{\rm YM}\rightarrow 0$. Notice that there
are no such static zero modes for fermionic components since they would not
satisfy the antiperiodic boundary conditions.
\footnote{This is a difference between our computation at finite temperature 
and computations of the Witten index for the matrix theory where
fermions obey periodic boundary conditions.}
This is an important
difference from the  zero temperature case and a manifestation of the fact
that supersymmetry is explicitly broken at non-zero temperature.

In the following, we will construct
an effective action for these coordinates by integrating the
off-diagonal gauge fields, the fermionic variables and the ghosts,
\begin{equation}
S_{\rm eff}[ \vec a^\alpha]\equiv -\ln \int[d a^\alpha_0]
\prod_{\alpha\neq\beta}[dA^{\alpha\beta}_\mu][d\theta][d{\rm ghost}]\exp\left(
-S_{\rm YM}-S_{\rm gf}-S_{\rm gh}\right).
\end{equation}
Generally, this integration can only be done in the a simultaneous
loop expansion and expansion in the number of derivatives of the
coordinates $\vec a^\alpha$.  Such an expansion is accurate in the
limit where $\left| \vec a^\alpha-\vec a^\beta\right|$ are large for
each pair of D0-branes and where the velocities are small.  Since
these variables are periodic in Euclidean time, small velocities are
only possible at low temperatures.  The remaining dynamical problem
then defines the statistical mechanics of a gas of D0-branes,
\begin{equation}
Z_{\rm YM}=\int\prod_{\tau,\alpha}
[d \vec a^\alpha(\tau)]\exp\left( -S_{\rm eff}[\vec a^\alpha]
\right).
\label{effec}
\end{equation}
We expect that the zero temperature limit of $S_{\rm eff}$ reduces to
(\ref{T=0}).  

We shall find several subtleties with this formulation.  If the
effective D0-brane action is to reproduce the results of a string
theoretical computation, the integration over $a_0^\alpha$ must be
performed in both cases.   

The effective action is a symmetric functional of the position
variables $\vec a^\alpha(\tau)$.  Only the configuration of these
coordinates needs to be periodic.  Therefore the individual position
should be periodic up to a permutation. The variables in the path
integral (\ref{effec}) should therefore be periodic up to a
permutation and the integral should be summed over the permutations.

\section{One loop computation}

We will compute the effective action $S_{\rm eff}$ in a simultaneous 
expansion in the number of loops and in powers of time derivatives 
of the D0-brane positions.

We decompose the gauge field into diagonal and off-diagonal parts,
\begin{equation}
A_\mu^{\alpha\beta}=a_\mu^{\alpha}\delta^{\alpha\beta}+g_{\rm YM}
\bar A^{\alpha\beta}_\mu
\end{equation}
where $\bar A_{\mu}^{\alpha\alpha}=0$ so that the curvature is
\begin{equation}
F_{\mu\nu}^{\alpha\beta}=\delta^{\alpha\beta}f^\alpha_{\mu\nu}+
g_{\rm YM}D_\mu^{\alpha\beta}
\bar A^{\alpha\beta}_\nu-g_{\rm YM}D_\nu^{\alpha\beta}
\bar A_\mu^{\alpha\beta}- ig^2_{\rm YM}
\left[ \bar A_\mu, \bar A_\nu \right]^{\alpha\beta}
\end{equation}
where
\begin{equation}
f_{\mu\nu}^\alpha= \partial_\mu a^\alpha_\nu-\partial_\nu a^\alpha_\mu
\end{equation}
and
\begin{equation}
D_\mu^{\alpha\beta}=\partial_\mu-i\left(a^\alpha_\mu-a^\beta_\mu\right).
\end{equation}
In the Yang-Mills term in the action, we keep all orders of the
diagonal parts of the gauge field and expand up to second order in the
off-diagonal components,
\begin{equation}
\frac{1}{g_{\rm YM}^2}{\rm TR}\left( F_{\mu\nu}^2 
\right)=\sum_\alpha \frac{1}{g_{\rm YM}^2}\left(f^\alpha_{\mu\nu}\right)^2
+2\sum_{\alpha\beta}\bar A_{\mu}^{\beta\alpha}\left(
\delta_{\mu\nu}{\buildrel \leftarrow\over D}_\lambda^{\beta\alpha}
\vec D_\lambda^{\alpha\beta}-{\buildrel \leftarrow\over
D}_\mu^{\beta\alpha}\vec D_\nu^{\alpha\beta}+2i\left(
f^\alpha_{\mu\nu}- f^\beta_{\mu\nu} \right)\right)\bar
A_\nu^{\alpha\beta}+\ldots
\end{equation}
We will fix the gauge
\begin{equation}
D_\mu^{\alpha\beta}\bar A_\mu^{\alpha\beta}=0.
\end{equation} 
This entails adding the Fadeev-Popov ghost term to the action
\begin{equation}
S_{\rm gh}=\int\sum_{\alpha\beta} \left\{\bar c^{\alpha\beta}\left( 
-D^{\alpha\beta}_\mu\right)^2c^{\beta\alpha}+i
g_{\rm YM}\bar c^{\beta\alpha}D_\mu^{\alpha\beta}
\left[\bar A_\mu,c\right]\right\}
\end{equation}
There is a residual gauge invariance under the abelian transformation,
\begin{eqnarray}
\bar A^{\alpha\beta}_\mu\rightarrow \bar A^{\alpha\beta}_\mu 
e^{i(\chi^\alpha -\chi^\beta)},
\nonumber \\
a_\mu^\alpha\rightarrow a_\mu^\alpha+\partial_\mu\chi^\alpha.
\end{eqnarray}
We shall use this gauge freedom to set the additional condition
\begin{equation}
\partial_0 a^{\alpha}_0=0
\end{equation}
and to fix the constant%
\footnote{These constants are related to the eigenvalues of the
holonomy 
$$
{\rm P} \exp{\left(i \int_0^\beta d\tau A_0(\tau)\right)}= 
\Omega^\dagger\;{\rm diag}\; \left(
e^{i\beta a_0^1},\ldots,e^{i\beta a_0^{\rm N}}\right) \Omega
$$
known as the Polyakov loop winding along the compact Euclidean time.
It can not be made trivial by the gauge transformation if $T\neq 0$.}
\begin{equation}
-\pi/\beta < a_0^\alpha\leq\pi/\beta.
\end{equation}

The ghost for this gauge fixing condition decouples.

Keeping terms up to quadratic order in $\bar A, c,\bar c, \theta$, the
action is
\begin{eqnarray}
S=\int\left\{\frac{1}{4g_{\rm YM}^2}\sum_\alpha \left( 
f^\alpha_{\mu\nu}\right)^2 +\frac{1}{2}
\sum_{\alpha\beta}\bar A_{\mu}^{\beta\alpha}
\left( -\delta_{\mu\nu} D_\lambda^{\beta\alpha} 
D_\lambda^{\alpha\beta}+D_\mu^{\beta\alpha}
D_\nu^{\alpha\beta}+2i\left( f^\alpha_{\mu\nu}- f^\beta_{\mu\nu}
\right)\right)\bar A_\nu^{\alpha\beta}
\right.\nonumber\\
\left.
+\sum_{\alpha\beta}\bar
c^{\beta\alpha}(D_\mu^{\alpha\beta})^2c^{\alpha\beta}
+\frac{i}{2}\theta^{\beta\alpha}\gamma_\mu D_\mu^{\alpha\beta}
\theta^{\alpha\beta}
\right\}.
\end{eqnarray}
The effective action obtained by integrating over $\bar A,\bar
c,c,\theta$ is
\begin{eqnarray}
S_{\rm eff}=\int\sum_\alpha\frac{1}{4g_{\rm YM}^2}(f^\alpha_{\mu\nu})^2
+\sum_{\alpha\neq\beta}\left\{ \frac{1}{2}{\rm TR}\ln\left( -\delta_{\mu\nu}(D^{\alpha\beta}_\mu)^2 +2i(f_{\mu\nu}^\alpha-f_{\mu\nu}^\beta)\right)
\right. \nonumber \\  \left.
-{\rm TR}\ln\left( -(D_\mu^{\alpha\beta})^2\right)-\frac{1}{2}{\rm TR}\ln
\left( i\gamma_\mu D^{\alpha\beta}_\mu\right) \right\}.
\label{seff}
\end{eqnarray}

\subsection{Leading order in time derivatives}

We will evaluate the determinants on the right-hand-side of
(\ref{seff}) in an expansion in powers of the derivatives of $\vec
a(\tau)$.  The leading order term can be found by setting $\vec a={\rm
const.}$. In this case, $f^\alpha_{\mu\nu}=0$ and 
(here we retain the tree-level term with time derivatives)
\begin{equation}
S_{\rm eff}=8\sum_{\alpha <\beta}\left\{{\rm
TR}_B\ln\left( -(D^{\alpha\beta}_\mu)^2 \right) -{\rm TR}_F\ln\left(
-(D^{\alpha\beta}_\mu)^2 \right)\right\}
\end{equation}
where the subscript $B$ denotes contributions from the gauge fields
and ghosts, whereas $F$ denotes those from the adjoint fermions.  The
determinants should be evaluated with periodic boundary conditions for
bosons and anti-periodic boundary conditions for fermions.  (Note
that, because of supersymmetry, if both bosons and fermions had
identical boundary conditions the determinants would cancel.  This
would give the well-known result that the lowest energy state is a BPS
state whose energy does not depend on the relative separation of the
D0-branes.)  
The boundary conditions are taken into account by introducing Matsubara 
frequencies, so that
\begin{equation}
e^{-S_{\rm eff}}=e^{-S_{0}} \beta^{\rm N}
\int \frac{da^\alpha_0}{2\pi} \prod_{\alpha<\beta}
\prod_{n=-\infty}^\infty
\left( \frac{ \left(\frac{2\pi n}{\beta}
+\frac{\pi}{\beta}+a_0^\alpha-a_0^\beta\right)^2+\vert\vec 
a_\alpha-\vec a_\beta\vert^2}{\left(  
\frac{2\pi n}{\beta}+a_0^\alpha-a_0^\beta\right)^2 +\vert\vec 
a_\alpha-\vec a_\beta\vert^2}
\right)^8
\end{equation}
Using the formula
\begin{equation}
\prod_{n=-\infty}^\infty
\left( \frac{2\pi n}{\beta}+\omega\right)=\sin\left( \frac{\beta\omega}
{2}\right)
\label{21}
\end{equation}
we obtain the result
\begin{equation}
e^{-S_{\rm eff}}=e^{-S_{0}} \beta^{\rm N}
\int \frac{da^\alpha_0}{2\pi} \prod_{\alpha<\beta}
\left(  \frac{ \cosh \beta\vert\vec a^\alpha-\vec a^\beta\vert+
\cos \beta \left(a_0^\alpha- a_0^\beta\right)}
{\cosh\beta\vert\vec a^\alpha-\vec a^\beta\vert - 
\cos\beta \left( a_0^\alpha-a_0^\beta\right) }\right)^8
\label{23}
\end{equation}
In order to find the effective action for $\vec a^\alpha$, 
it is now necessary to 
integrate the temporal gauge fields $a_0^\alpha$ over the domain 
$(-\pi/\beta,\pi/\beta]$.  This integration implements the projection 
onto the gauge invariant eigenstates of the matrix theory Hamiltonian.

In the case where there is a single pair of D0-branes, N=2, the integration
over $a_0^\alpha$ in (\ref{23}) can be done explicitly to obtain
the effective action
\begin{equation}
S_{\rm eff}=\int_0^\beta d\tau\left\{
\sum_1^2\frac{(\dot{\vec a}{}^\alpha)^2}{2g_{\rm YM}^2} 
 -\frac{1}{\beta}\ln\left(
\frac{ P(z) }{  (1-z^2)^{15} }
\right)\right\}
\label{effac}
\end{equation}
where 
\begin{eqnarray}
P(z)= 1+241z^2+12649z^4+254009z^6+2434901z^8+12456773z^{10}
+36119181z^{12}\nonumber \\*+61178589z^{14}+6191459z^{16}
+36109171z^{18}
+12462779z^{20}+2432171z^{22}\nonumber\\* 
+254919z^{24}+12439z^{26}+271z^{28}
-z^{30},
\end{eqnarray}
$z=\exp(-\beta\vert {\vec a}^1-{\vec a}^2\vert)$ and we have
included the tree level term, which gives the non-relativistic kinetic
energies of the D0-branes.

The effective action has the
low temperature expansion
\begin{eqnarray}
S_{\rm eff}=\int_0^\beta d\tau\left( \frac{1}{2 g_{\rm YM}^2}\sum_\alpha 
\left(\dot{\vec a}^\alpha\right)^2-\frac{1}{\beta}\left( 256 e^{-2\beta\vert 
\vec a^1-\vec a^2\vert}-16384 e^{-4\beta\vert \vec a^1-a^2
\vert} \right.\right. \nonumber \\*   \left.\left.
+\frac{5614336}{3}e^{-6\beta\vert \vec a^1-\vec a^2\vert}+
\ldots\right)\right)
\label{finalSeff}
\end{eqnarray}
We shall compare in the following section this result with
the superstring computation of the effective interaction between D0-branes. 

\section{String theoretical interactions}

The effective interactions of D0-branes 
in superstring theory is given by computing the annulus diagram 
shown in fig.~\ref{f:annulus}. 
This was done in ref.~\cite{green} (and in~\cite{VM96} for Dp-branes).  
The result of summing over physical (GSO projected) superstring states gives 
the free energy
\begin{equation}
F[L,\beta,\nu]=\frac{8}{\sqrt{\pi \alpha'}}\int_0^\infty \frac{dl}{l^{3/2}}\,
e^{ -L^2l/4\pi^2\alpha'}
\Theta_2 \left(\nu \left \vert \frac{i\beta^2}{\pi \alpha' l} \right. \right)
\prod_{n=1}^\infty\left( \frac{1+e^{-nl}}{1-e^{-nl}}\right)^8
\label{GreenF}
\end{equation}
where
\begin{equation} 
\Theta_2 \left(\nu \left \vert iz \right. \right)
=\sum_{q=-\infty}^\infty
e^{ -\pi z(2q+1)^2/4+i\pi(2q+1)\nu},
\end{equation}
$L$ is the brane separation 
and $\nu$ is a parameter
which weights the winding numbers of strings around the periodic time 
direction. An extra factor of 2 accounting for the exchange of the
two ends of the superstring~\cite{Pol96} ending on each of the two 
D0-branes is inserted in~(\ref{GreenF}). 

The product in the integrand represents the sum over string states, 
with requisite degeneracies,
\begin{equation}
8\prod_{n=1}^\infty\left( \frac{1+e^{-nl}}{1-e^{-nl}}\right)^8
=\sum_{N=0}^{\infty} d_N e^{-Nl}
\label{degeneracy}
\end{equation}
where $d_N$ is the degeneracy of the 
either superstring state at level $N$.  For 
the lowest few levels, $d_0=8$ and $E_0=L/2\pi\alpha'$.

Inserting (\ref{degeneracy}) in (\ref{GreenF}) and integrating over $l$,
the free energy has the form
\begin{equation}
F(\beta,L,\nu)=\frac{2}{\beta}\sum_{N=0}^\infty d_N~
\ln\left| \frac{ 1-e^{-\beta E_N+i\pi\nu}}{ 1+e^{-\beta E_N+i\pi\nu} }
\right|
\label{Fsuperstring}
\end{equation}
where the string energies are given by the formula
\begin{equation}
E_N=\frac{L}{2\pi\alpha'}\sqrt{ 1+\frac{4\pi^2\alpha'N}{L^2}}.
\label{Esuperstring}
\end{equation}
This results in the partition function
\begin{equation}
Z_{\rm str}(\beta,L,\nu)\equiv e^{-\beta F}= \prod_{N=0}^\infty \left|
\frac{1+e^{-\beta E_N+i\pi \nu}}{1-e^{-\beta E_N+i\pi \nu}}
\right|^{2d_N}.
\label{Zsuperstring}
\end{equation}
The physical meaning of the last formula is obvious:
the partition function equals the ratio of the Fermi and
Bose distributions with the power being twice the degeneracy of states
and $i\nu$ playing the role of a chemical potential. The factor of 2
in the exponent $2d_N$ in (\ref{Zsuperstring}) 
is due to the interchange of the superstring ends as is already
mentioned. It will provide the agreement with the matrix theory 
computation.

In order to compare with the Yang-Mills computation, we should first re-scale 
the coordinates so that the mass of the D0-brane appears in the kinetic term
as in (\ref{T=0}).  
The mass is given by the formula
\begin{equation}
M=\frac{1}{g_s\sqrt{\alpha'}}
\end{equation}
and the Yang-Mills coupling $g_{\rm YM}$ is related to the string coupling 
$g_s$ by the equation
\begin{equation}
g_{\rm YM}^2=\frac{g_s}{4\pi^2(\alpha')^{3/2}}.
\end{equation}
The physical coordinate of the $\alpha$-th D0-brane is identified with
\begin{equation}
\vec q^\alpha= 2\pi\alpha'\vec a^\alpha .
\label{qvsa}
\end{equation}
Taking N=2 in (\ref{23}) and identifying
$L=2\pi\alpha' \vert \vec a^1-\vec a^2 \vert$, we see that the 
integrand in (\ref{23}) coincides with (\ref{Zsuperstring})
truncated to the massless modes ($N=0$) provided 
$\nu=\beta(a_0^1-a_0^2)$.

It is clear that the integral over $a_0^\alpha$ is responsible for 
the ``mismatch'' of  the effective 
actions between the string theory computation 
and matrix theory computation of the free energy. 
In the string theory done in the spirit of ref.~\cite{green}, 
the parameter $\nu$ appears in the same place as 
$\beta(a_0^1-a_0^2)/\pi$ but is not integrated. It is associated
with the interaction of the ends of the open string with an Abelian
gauge field background ${A}_{\mu}(\tau,\vec x)$: 
\begin{equation}
S_{\rm int}=\int dx^\mu {A}_\mu.
\end{equation}
If the ends are separated by the distance $L$, e.g.\ along the
first spatial axis, then
\begin{equation}
\nu=\int_0^\beta d\tau \left( {A}_0(\tau,0,\ldots)-
{A}_0(\tau,L,\ldots)\right)
\end{equation}
since $\dot x_\mu (\tau)=(1,\vec 0)$ on the boundaries.
The matrix theory automatically takes into account the
integration over the background field while in the string theory
calculation of ref.~\cite{green} the background field is fixed.
This is just a reflection of the fact that matrix theory is 
an effective low-energy theory of $D$-branes, while the 
older string theory did not treat the boundaries as dynamical 
objects. However, it is interesting to notice how close some 
of the earlier string papers came to such a description simply 
by the requirement of consistency \cite{green1}. Further, 
in the context of matrix theory it is natural to take the 
exponential of the free energy (\ref{Fsuperstring}) as in
eq.\ (\ref{Zsuperstring}), and only integrete over $\nu$
afterwards, a procedure not entirely obvious in 
a string theory where the boundaries are not dynamical objects.
This will be discussed further in the next section.

An exact coincidence between the matrix theory and superstring results
is possible only when the higher stringy modes are suppressed.
Usually, the truncation of the string spectrum to get
Yang-Mills theory is valid for small $\alpha'$, that is when we are interested
in temperatures which are much smaller than $1/\sqrt{\alpha'}$.  In fact, the 
condition in our case  
is a little different than this once the length $L$ appears as a parameter in 
the spectrum (\ref{Esuperstring}).  
Then, the spectrum can be truncated at the first level only when
\begin{equation}
\frac{1}{\beta}\equiv k_BT\ll 
\sqrt{ \left(  \frac{L}{2\pi\alpha'}\right)^2+\frac{1}{\alpha'} }-\frac{L}
{2\pi\alpha'}
\label{trunk}
\end{equation}
which is the energy gap between the first two levels.
If the temperature is small, this condition is always satisfied unless
the length $L$ is not too large. In other words the truncation of the
spectrum to the lightest modes is valid for $\beta \gg L$
(or $TL\ll 1$).

It is also interesting to discuss what happens in the opposite
limit $L \gg \beta$ where the interaction between D0-branes
is mediated by the lightest closed string modes.
The superstring free energy can be evaluated in this limit by the
standard modular transformation which relates the annulus diagram
for an open string with a cylinder diagram for a closed string.
Introducing the new integration variable $s=2\pi^2/l$, we 
rewrite (\ref{GreenF}) as~\cite{green}
\begin{equation}
 F[L,\beta,\nu]= \frac{8 \pi^4}{\sqrt{2 \pi \alpha'}} 
\int_0^\infty \frac{ds}{s^{9/2}}\;
e^{s}\,e^{-L^2/2s\alpha'}
\Theta_2 \left(\nu \left \vert \frac{i\beta^2 s}{2\pi^3 \alpha' } 
\right. \right)
\prod_{n=1}^\infty\left( \frac{1-e^{-(2n+1)s}}{1-e^{-2ns}}\right)^8.
\label{viathetas}
\end{equation}
In the limit where the brane separation is large the integration over $s$ is 
concentrated in the region of large $s\sim L^{2}$.
Substituting the large-$z$ asymptotics
\begin{equation}
\Theta_2 \left(\nu \left \vert i z  \right. \right)
\rightarrow
2\cos{(\pi\nu)}e^{-\pi z/4}
\end{equation}
and evaluating the saddle-point integral, we get
\begin{equation}
 F[L,\beta,\nu] \propto \cos{(\pi\nu)} 
\frac{\left(\beta^2-8\pi^2\alpha'\right)^{3/2}}{L^4}
e^{-L\sqrt{\beta^2-8\pi^2\alpha'}/2\pi \alpha'}.
\end{equation}

Exponentiating and integrating over $\nu$, we have
\begin{equation}
\int_{-1}^{1} {d\nu} Z_{\rm str}(\beta,L,\nu) 
\propto \frac{\beta^2 \left(\beta^2-8\pi^2\alpha'\right)^{3}}{L^8}
e^{-L\sqrt{\beta^2-8\pi^2\alpha'}/\pi \alpha'}.
\end{equation}
Taking into account (\ref{qvsa}) the exponent at the low temperatures is
the same as in (\ref{finalSeff}) but the pre-exponential differs.
The dependence of the pre-exponential on $L$ in the superstring case 
emerges because the splitting between energy states in
(\ref{Esuperstring}) is of order $1/L$ and the truncation condition
(\ref{trunk}) is no longer satisfied when $L$ is large.
Higher stringy modes are then not separated 
by a gap and the continuum spectrum 
results in the $L$-dependence of the preexponential.
As usual, the limits of $L\rightarrow\infty$ and $T\rightarrow 0$ are not 
interchangeable in the superstring theory.

\section{Discussion}

Our main results concern D0-brane dynamics at finite temperatures. 
We have computed the 1-loop effective action for the interaction
of static D0-branes in the matrix theory at finite temperature and
compared it with the analogous superstring computation.
We have seen that an extra integration over the eigenvalues
of the holonomy along the compactified Euclidean time is present
in the matrix theory. The two computations agrees in the 
low temperature limit provided the superstring thermal partition
function is integrated over the Abelian gauge fields $a_0$'s living on
D0-branes. 

The integration over $a_0$'s is of course natural in the context 
of the Yang-Mills theory, where it expresses that only 
gauge-invariant states should contribute to ${\rm TR}\,e^{-\beta H}$.
But it is also natural from the point
of view of the D0-brane physics. It can be seen as follows. 
Suppose we make a T-duality transformation, which interchanges the Neumann
and Dirichlet boundary conditions,
along the compactified Euclidean time direction. Then $a_0$'s become
the coordinates of D-instantons on the dual circle. The integration
over $a_0$'s becomes now the integration over the positions of D-instantons.
The partition function should involve such an integration over
the collective coordinates and since they are collective coordinates
the integration appears in front of the exponential of the effective 
action, not in the action itself. Viewed in terms of D0-branes and 
open strings, we have a gas of D0-branes with open strings between 
them. The individual strings might have a winding number $q$ (more 
precisely $2q+1$ in the case of superstrings), describing the 
winding around the finite-temperature space-time cylinder. The energy 
of such states are $\propto \beta q/2\pi \alpha'$. However, the 
$q$'s satisfy $\sum q=0$ as a result of the integration over $a_0$.
Physically this constraint is most easily understood  by going to 
the closed string channel where we have closed string boundary state 
on the dual circle with radius $\tilde{\beta} = 4\pi^2 \alpha'/\beta$
localized at the point $(\nu \tilde \beta, \vec q ) $. Passing to the momentum
representation, we write
\begin{equation}
\Big\vert B,\vec q, \nu \Big\rangle =
\sum_{q=-\infty}^\infty e^{-2i\pi\nu q} \, \Big\vert B,\vec q, 
p_0=2\pi q /\tilde{\beta} \Big\rangle .
\end{equation}
Here the temporal momentum is quantized as 
$p_0=2\pi q /\tilde{\beta}=q\beta/2\pi\alpha'$, which 
lead to the same energy as the above mentioned open string states.
In this representation $\sum q=0$ simply expresses momentum conservation
in the thermal direction.

The effective static potential between two D0-branes emerges because 
supersymmetry
is broken by finite temperature. This effect of breaking supersymmetry
is somewhat analogous to the velocity effects at zero temperature
where the matrix theory and superstring computations agree to
the leading order of the velocity expansion~\cite{dkps}.
We have thus shown that the leading term in a low temperature expansion 
is correctly reproduced by the matrix theory.
The discrepancy between the matrix theory and superstring computations, 
which we have observed in the limit of large distances $L T\gg 1$,
does not contradict this statement since temperature the limits of large 
distances and small temperatures are not interchangeable.

An interesting feature of the effective static potential between
D0-branes is that it is logarithmic and attractive at short distances.
In the matrix theory, the singularity of the computed 1-loop potential
occurs when the distance between the D0-branes vanishes and the 
SU(N) symmetry which is broken by finite distances is restored.
The integration over the off-diagonal components can no longer be treated
in the 1-loop approximation! 
In the superstring theory, the singularity is exactly the same as in the
matrix theory since it is determined only by the massless bosonic modes
(the NS sector in the superstring theory). Its origin is {\em not}\/ due 
to the presence of massless photon states in the spectrum.
Putting $\nu_1=\nu_2$ in the above D-instanton picture on the dual
temporal circle, we see that the mass of the lowest states,
associated with the winding numbers $2q+1=\pm 1$ is 
$\tilde \beta{} /2\pi\alpha'$. The divergence at $L=0$ emerges, in
this picture, after summing over all the open string states since
no single state has such a divergence. It shows up only
at finite temperature where the winding number $q$ exists.

It is important to notice that the computed partition functions take
into account only thermal fluctuations of superstring stretched
between D0-branes but not the fluctuations of D0-branes themselves.  
To calculate the thermal partition function of D0-branes,
a further path integration over their periodic trajectories $\vec a(\tau)$ 
is to be performed as in (\ref{effec}). One might think that 
classical statistics is applicable to this problem since the D0-branes
are very heavy as $g^2_{\rm YM}\rightarrow0$ so that one could restrict
himself by the static approximation. 
This is however not the case
due to the singularity of the effective static potential at small distances.
The integral over the D0-brane positions $\vec{a}$'s is divergent
when the two positions coincide.

However, this singularity is only in the classical partition function.  
The path integral over the periodic trajectories $\vec a(\tau)$ 
that we actually have to do can not diverge since 
the 2-body quantum mechanical problem has a well-defined spectrum. 
 The path integral can then be evaluated as
$\sum_n \exp(-\beta E_n)$
where $E_n$ are in the spectrum of the operator
$ H=P^2/M+V_{\rm eff}$. There certainly should not be the bound state energy  
eigenvalue at negative infinity for this
quantum mechanical problem which implies the convergence of the
path integral. These issues which are related to thermodynamics of
D0-branes will be considered in a separate publication. 

Let us finally discuss when the 1-loop appoximation
that we have done is applicable.
The loop expansion in Yang-Mills theory computation 
is valid only in the limit where
$$
g_{\rm YM}^2/\vert\vec a\vert^3\sim
g_s \left(\frac{ \sqrt{\alpha'}}{L}\right)^3
$$ 
is small.  This is due to the fact that the distance $L$ plays the role of a 
Higgs mass which cuts off the infrared divergences of the loop expansion in 
the 0+1~-dimensional gauge theory.
Thus, the perturbative Yang-Mills theory computation is good when 
\begin{equation}
g_s^{1/3}\sqrt{\alpha'} \ll L.
\end{equation}   
This can be satisfied if either the string coupling is small or if the 
D0-brane 
separation is large compared with the string length scale. In the latter case,
the truncation of the spectrum to the lightest modes is still valid when
\begin{equation}
k_BT\ll \frac{1}{L} \ll\frac{1}{g_s^{1/3}\sqrt{\alpha'}}
\end{equation} 
Note that the first inequality which is independent of both the string 
scale and the string coupling is the one already discussed in
the previous section. In this case, the temperature must be less than 
the inverse distance between D0-branes.   
In the case where the string coupling $g_s$ is small, the criterion for 
validity of truncation of  the spectrum becomes
\begin{equation}
k_BT<1/\sqrt{\alpha'}
\end{equation}
that is the usual one.

\section*{Acknowlegments}

We are gratefull to P.~Di~Vecchia, G.\ Ferretti, A.~Gorsky, M.~Green, 
M.~Krogh,  
N.~Nekrasov, N.~Ohta, P.~Olesen, L.~Thorlacius and K.~Zarembo 
for very useful discussions. 
J.A. and Y.M. acknowledge the support 
from MaPhySto financed by the Danish National Research Foundation
and from INTAS under the grant 96--0524.  
The work by Y.M. is supported in part by the grants
CRDF 96--RP1--253 and RFFI 97--02--17927.

\end{document}